\documentstyle[psfig]{mn} 

\title[Chandra Observations of M82]{Chandra High-Resolution
Camera Observations of the Luminous X-Ray Source in the
Starburst Galaxy M82} 

\author[P. Kaaret et al.]{P.~Kaaret$^1$, A.H.~Prestwich$^1$,
A.~Zezas$^1$, S.S.~Murray$^1$,  D.-W.~Kim$^1$, \newauthor
R.E.~Kilgard$^1$, E.M.~Schlegel$^1$ and M.J.~Ward$^2$ \\
$^1$Harvard-Smithsonian Center for Astrophysics, 60 Garden
St., Cambridge, MA 02138, USA\\ $^2$Department of Physics and
Astronomy, University of Leicester, Leicester LE1 7RH, UK}

\date{Accepted 2000. Received 2000 ; in original form 2000
August 8}

\pagerange{\pageref{firstpage}--\pageref{lastpage}}
\pubyear{2000}

\begin{document}

\maketitle

\label{firstpage}

\begin{abstract}

We have analyzed Chandra High Resolution Camera observations
of the starburst galaxy M82, concentrating on the most
luminous x-ray source.  We find a position for the source of
$\rm R.A. = 09^h 55^m 50^s.2, decl. = +69^{\circ} 40\arcmin
46\arcsec.7$ (J2000) with a $1\sigma$ radial error of
$0.7\arcsec$.  The accurate x-ray position shows that the
luminous source is not at the dynamical centre of M82 nor
coincident with any suggested radio AGN candidate.  The
source is highly variable between observations, which
suggests that the source is a compact object and not a
supernova or remnant. There is no significant short term
variability within the observations.  Dynamical friction and
the off-center position place an upper bound of $10^{5} -
10^{6} \, M_{\odot}$ on the mass of the object, depending on
its age.  The x-ray luminosity suggests a compact object mass
of at least $500 \, M_{\odot}$.  Thus, the luminous source in
M82 may represent a new class of compact object with a mass
intermediate between those of stellar mass black hole
candidates and supermassive black holes.

\end{abstract}

\begin{keywords} black hole physics -- galaxies: individual:
M82 -- galaxies: starburst -- galaxies: stellar content --
X-rays: galaxies \end{keywords}

\section{Introduction}

One of the most enigmatic results to emerge from X-ray
population studies of spiral and other luminous star forming
galaxies is the discovery of unresolved X-ray sources which
appear to have luminosities factors of 10 to 100's times the
Eddington luminosity for a neutron star (e.g.\ Roberts \&
Warwick 2000; Zezas, Georgantopoulos, \& Ward 1999;  Wang,
Immler, \& Pietsch 1999; Colbert \& Mushotzky 1999; Fabbiano,
Schweizer, \& Mackie 1997; Marston et al.\ 1995; for a review
of early results see Fabbiano 1989).  The origin of such
sources is controversial.  Some are located near the
dynamical centre of the host galaxy, and hence may be low
luminosity AGN.  However, many are well outside the central
regions of the galaxies and require an alternative
explanation.  Some of these highly luminous x-ray sources may
be very luminous supernova remnants exploding into a dense
interstellar medium \cite{fabian96}, or they may be accretion
powered binary sources, in which case they are excellent
black hole candidates with masses near or above $10
\,M_{\odot}$ \cite{makishima00}.  Deciding between these
various alternatives has been complicated by the limited
spatial resolution of pre-Chandra X-ray missions.

One of the most extreme and controversial examples of a
highly luminous x-ray source in a nearby galaxy is the bright
X-ray source that dominates the central region of the nearby
starburst galaxy M82.  Previous Einstein, ROSAT, and ASCA
observations have shown that this source is variable and is
close to the centre of M82 \cite{watson84,collura94,ptak99}. 
It has been interpreted as a low luminosity AGN
\cite{tsuru97}, a highly x-ray luminous supernova
\cite{stevens99}, and an accreting black hole with a mass in
excess of $460 M_{\odot}$ \cite{ptak99}.  In this paper we
discuss early Chandra observations of M82 made using the High
Resolution Camera (HRC; Murray et al.\ 1997).  The central
x-ray `source' in M82 is resolved into several sources in the
HRC observations.  We present an analysis of the brightest
Chandra source.  Our results suggest that this source may be
a black hole with a mass intermediate between stellar-mass
Galactic x-ray binaries and supermassive black holes.  We
describe the observations in \S~2, our analysis in \S~3, and
conclude in \S~4.

\section{Observations}


\begin{figure*}
\centerline{\psfig{file=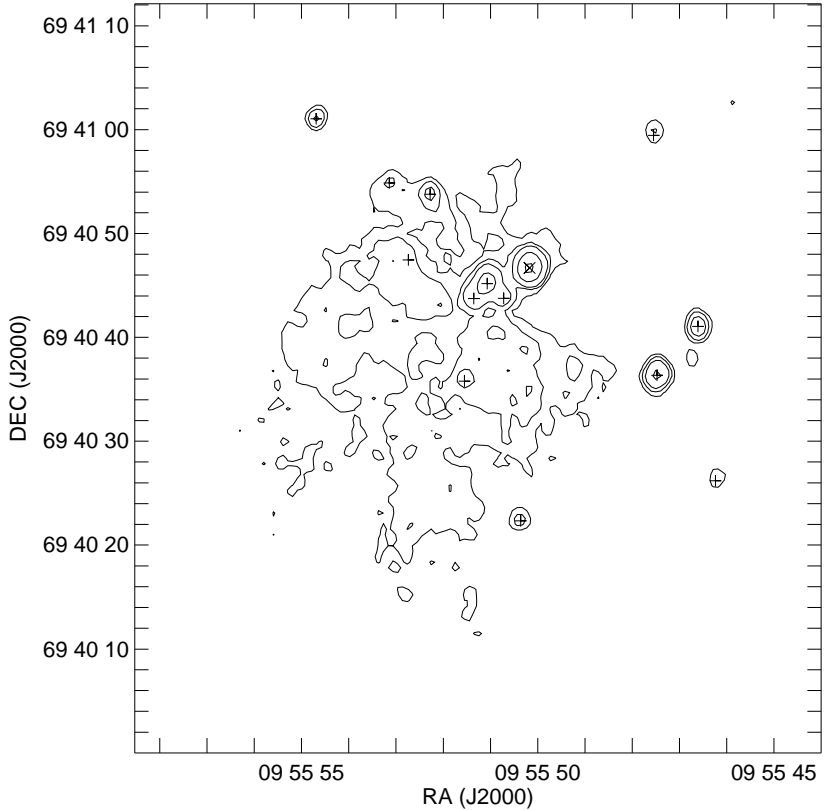,width=4.5in}}  \caption{The
central region of M82 from the October 1999 observation.  The
contours were calculated using a counts map with $0.53\arcsec$
pixels smoothed with a Gaussian with FWHM = $1.06\arcsec$. The
contours indicate 2.5, 5, 10, 40, and 160 counts per pixel in
the smoothed map.  The position of X41.4+60 is marked with an
`X'.  Crosses indicate positions of other sources.}
\label{m82image} \end{figure*}

M82 was observed with the Chandra X-Ray Observatory (CXO;
Weisskopf 1988) using the High Resolution Camera (HRC; Murray
et al.\ 1997) and the High-Resolution Mirror Assembly (HRMA;
van Speybroeck et al.\ 1997) on 1999 Oct 28 04:24 UT to 14:48
UT for an exposure of 36~ks and on 2000 Jan 20 14:51 UT to
20:25 UT for an exposure of 18~ks.  The HRC is a microchannel
plate imager having very good spatial and time resolution,
but essentially no energy resolution.  Each photon detected
by the HRC is time and position tagged, making possible
timing studies of individual sources in crowded fields.  The
HRC contains a wiring error, discovered after launch (Murray
et al.\ 2000), which induces a  3--4~ms error in the event
time tags for this observation.  As we restrict ourselves to
frequencies below 1~Hz, this error has no effect on the
analysis presented below.  The HRC position tags have a
precision of $0.132\arcsec$, referred to as `one pixel'. 
This resolution oversamples the Chandra point spread function
(PSF) which has a half-power diameter of $0.76\arcsec$
(Jerius et al.\ 2000).  We used a 15.6 pixel radius to
extract source light curves.

We applied aspect to X-ray events from the HRC and filtered
the data using event screening techniques (Murray et al.\
2000) to eliminate `ghost' events produced by the HRC
electronics.  An image for each observation was generated
from the filtered event lists, see Fig.~\ref{m82image}.  We
used the  standard Chandra software routine wavdetect to
search for and determine the position of point sources  (CIAO
V1.1 Software Tools Manual).  We found several sources in
each observation including both transients and persistent
sources.  Here, we concentrate on the brightest source found.
The other sources, including spectroscopy from observations
with the Chandra Advanced CCD Imaging Spectrometer (ACIS;
Bautz et al.\ 1998), will be described in a forth-coming
paper \cite{m82acis}.

\section{Results}

The brightest source in both observations is at a location of
$\rm R.A. = 09^h 55^m 50^s.2, decl. = +69^{\circ} 40\arcmin
46\arcsec.7$ (J2000).  Following the convention of naming
sources in M82 via their offset from $\rm R.A. = 09^h 51^m
00^s, decl. = +69^{\circ} 54\arcmin 00\arcsec$ (B1950), we
refer to this source as X41.4+60 in the remainder of the
paper.  For wider use, we also denote the source as
CXOU~J095550.2+694047.  The position uncertainty is dominated
by the accuracy of the aspect reconstruction which we take to
have a $1\sigma$ radial error of $0.7\arcsec$ (Aldcroft et
al.\ 2000).  

The source lies $9\arcsec$ from the kinematic centre of M82
\cite{weliachew84}, $12\arcsec$ from the $2.2 \, \mu m$ peak
\cite{rieke80}, $4\arcsec$ from the very luminous supernova
remnant 41.95+57.5 \cite{kronberg75,wills97}, and $13\arcsec$
from the suggested AGN candidate 44.01+59.6
\cite{wills97,seaquist97,wills99}. The radio source
41.31+59.6, which is likely a compact supernova remnant
\cite{muxlow94,allen98}, lies near the edge of the error
circle.  The highly variable radio source 41.5+59.7
\cite{kronberg85} lies within the error circle.  This radio
source was bright in one observation in 1981 but not detected
one year later or subsequently, and has been interpreted as
due to a supernova \cite{kronberg00}.  However, as only one
detection is available and unique identification is not
possible based on the radio spectral index alone; the source
may belong to a different class of radio transient
\cite{muxlow94}.  If the 1981 radio event was a supernova, it
is likely unrelated to the x-ray source as a bright x-ray
source was detected at this position in 1979 with the
Einstein High Resolution Imager \cite{watson84}. The 408~MHz
radio flux at the position of X41.4+60 is below 2~mJy
\cite{wills97}.  The 6~cm radio flux is below 2~mJy except
during 1981 \cite{kronberg00}.

X41.4+60 is a highly variable x-ray source.  In the first
observation, the HRC count rate from the source is 0.07~c/s,
while in the second it is 0.52~c/s -- a factor of 7 brighter.
In the first observation, X41.4+60 accounts for roughly 40\%
of the counts within $6\arcsec$, i.e.\ comparable to the
resolution of the ROSAT HRI, of the source position and only
8\% of the counts within $4\arcmin$, i.e.\ comparable to the
resolution of the ASCA SIS.  In the second observation,
X41.4+60 accounts for more than 90\% of the counts within
$6\arcsec$ and 40\% within $4\arcmin$.  Thus, even in the
brightest states of X41.4+60, ASCA spectra of the source are
significantly contaminated by flux from other point sources
and diffuse emission.  The time scale of the variability
places an upper limit on the size of the emitting region of
0.08~pc.  

Using various spectral models consistent with the spectrum of
this source extracted from a Chandra ACIS observation of M82
\cite{m82acis}, we estimate that 1~c/s in the HRC corresponds
to an observed flux of $0.9-1.4 \times 10^{-10} \rm \, erg \,
cm^{-2} \, s^{-1}$ in the 0.2--10~keV band.  The range in the
conversion factor is due to uncertainty in the ACIS spectral
fits.   The source flux in the ACIS observation corresponds
to an HRC rate of 0.03~c/s, so the true conversion factors
may differ from these values if the source spectrum varies
with flux.  The flux in the second observation is comparable
to the highest fluxes observed from the central source in M82
with ASCA \cite{ptak99}.  Taking a distance to M82 of
3.63~Mpc \cite{m82dist_hst}, the inferred isotropic source
luminosity from the absorbed flux in the two observations
would then be $1.0-1.5 \times 10^{40} \rm \, erg \, s^{-1}$
in the 0.2--10~keV band for the first observation and $7-11
\times 10^{40} \rm \, erg \, s^{-1}$ for the second. 
Correcting for absorption would increase these luminosities;
conversely, the true luminosity may be lower if the x-rays
are beamed.  These luminosities are near or above the highest
values found for non-nuclear sources in a ROSAT sample of
nearby galaxies \cite{roberts00}.


\begin{figure} \psfig{file=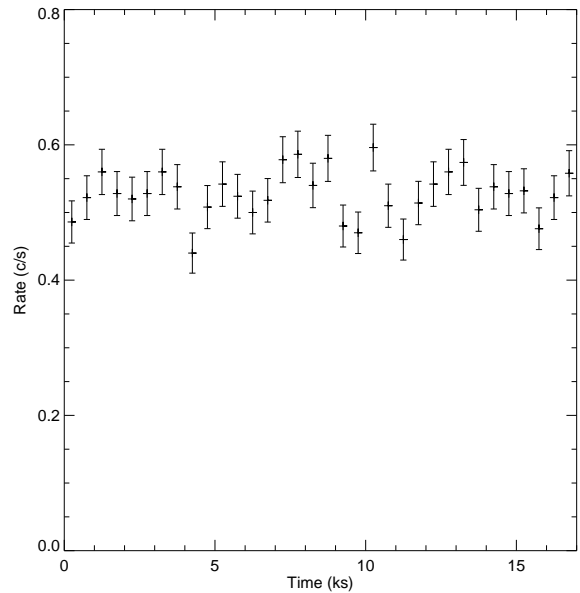,width=3.25in}
\caption{Light curve for X41.4+60 from the HRC observation in
January 2000.  The time bin size is 500~s.  Statistical error
bars are shown.} \label{lc} \end{figure}

For each observation, we extracted a light curve for
X41.4+60, see Fig.~\ref{lc}.  The source appears to have
roughly constant flux in both observations.  In particular,
the light curve for the January 2000 observation shows no
evidence of significant change in flux level over the
observation, so it is unlikely that the high flux represents
a flare of short duration.  In both observations, the power
spectra show no significant short term variability with the
power comparable to the Poisson noise limit over the
frequency range 0.0005-1~Hz.

We note that oscillations with a frequency near 600~s were
reported in a preprint of this manuscript.  In subsequent
analysis, these oscillations were found to be due to
instrumental effects related to the HRC response to bright
sources and the algorithms used to screen `ghost' events
produced by the HRC electronics.

\section{Discussion}

The luminosity of X41.4+60 in the January 2000 observation,
given the assumptions concerning the spectral shape and
isotropic emission noted above, corresponds to the Eddington
luminosity for a $500-900 \rm M_{\odot}$ object.  The strong
variability between observations argues against this
luminosity being due to an aggregate of sources.  The
variability also suggests that the source is a compact object
and not a supernova, although the possibility of a supernova
expanding into a highly non-uniform medium cannot be
excluded.  Soft gamma repeaters (SGRs) produce sufficient
flux; however, the longest, bright, so-called ``giant'',
outbursts from SGRs last only $\sim 300 \rm \, s$ over which
they show substantial decay \cite{hurley99}.  The fact that
X41.4+60 shows no evidence of decay or variability over
15~ks, see Fig.~\ref{lc}, argues against it being a SGR. 
Origin of the high luminosity and variability in an accreting
massive compact object appears plausible.

Dynamical friction will cause massive objects orbiting in the
stellar field surrounding a galactic nucleus to spiral into
the nucleus \cite{tremaine75}.  The life time, $t$, before
reaching the nucleus is related to mass of the object, $M$,
its distance from the nucleus, and the stellar velocity
dispersion.  From the position given above for X41.4+60 and
adopting a velocity dispersion for M82 of $100 \rm \, km \,
s^{-1}$ \cite{gaffney93}, a rough upper bound can be placed
on the mass of X41.4+60, $M \la 10^5 M_{\odot} (t/10^{10} {\,
\rm yr})^{-1}$.  If the object was formed during the initial
formation of M82 then $t \sim 10^{10} \rm yr$ hence $M \la
10^5 M_{\odot}$.  Higher masses are allowed if shorter life
times are assumed, e.g.\ $M \la 10^6 M_{\odot}$ for $t \sim
10^{9} \rm yr$.  This may be possible if the object was
formed recently, in a process likely to be distinct from that
for the formation of supermassive black holes in galactic
nuclei, or if the object was ejected from the nucleus in an
encounter with one or two equally or more massive black
holes.  If the object was formed recently and outside of the
nucleus, it is likely to be less massive than the super star
clusters found in M82 -- the most massive of which is $2
\times 10^{6} M_{\odot}$ \cite{smith00,oconnell95}.  Rapid
formation of a compact object with mass greater than $100
M_{\odot}$ in the collapse of a super star cluster appears
possible \cite{portegies99,taniguchi00}.

In conclusion, the accurate x-ray position determination for
the most luminous x-ray source in M82, made possible by the
high angular resolution of Chandra, excludes identification
with suggested radio AGN candidates \cite{wills97}.  The
strong variability of the source argues against the
possibility that it could arise from a supernova
\cite{ptak99,matsumoto99} and the high flux places a lower
bound on the compact object mass of $500 \, M_{\odot}$ if the
emission is isotropic.  The low radio flux at the x-ray
position and the displacement of the source from the
dynamical centre of M82 argue against X41.4+60 being a
supermassive black hole similar to that seen at the centre of
the Milky Way.  The most plausible explanation for the object
is that it is an accreting black hole with a mass of $500 -
10^{5} \, M_{\odot}$.  The Chandra data strengthen this
suggestion made previously on the basis of ASCA data by Ptak
\& Griffiths (1999) and also clearly establish that the
source is non-nuclear \cite{matsumoto99}.  In this case, it
would represent a new class of compact object
\cite{colbert99} with a mass intermediate between those of
stellar mass black hole candidates and supermassive black
holes found in the centres of galaxies.  Understanding the
formation of such an object \cite{portegies99,taniguchi00}
may provide insights into the formation of super-massive
black holes in galactic centers \cite{quinlan90,gebhardt00}.

\section*{Acknowledgments}

We thank Tom Aldcroft for information concerning the dither
and aspect solution and thank Doug Richstone, Andy Fabian,
Avi Loeb, Roger Blandford, Martin Weisskopf, Frank Primini,
Hironori Matsumoto, and Wallace Tucker for useful
discussions.  We gratefully acknowledge the efforts of the
Chandra team and support from NASA Chandra contract
NAS8-39073.  PK acknowledges partial support from NASA grant
NAG5-7405.


\label{lastpage}

\end{document}